# Low Symmetry Phase in (001) BiFeO$_3$ Epitaxial Constrained Thin Films

Guangyong Xu, H. Hiraka$^{(a)}$, and G. Shirane

*Physics Department, Brookhaven National Laboratory, Upton, New York 11973*

Jiefang Li, Junling Wang, and D. Viehland

*Department of Materials Science and Engineering, Virginia Tech, Blacksburg, VA 24061*

(Dated: January 20, 2005)

## Abstract

The lattice of (001)-oriented BiFeO$_3$ epitaxial thin film has been identified by synchrotron x-ray diffraction. By choosing proper scattering zones containing the fixed (001) reflection, we have shown that low-symmetry phases similar to a $M_A$ phase exist in the thin film at room temperature. These results demonstrate a change in phase stability from rhombohedral in bulk single crystals, to a modified monoclinic structure in epitaxial thin films.

PACS numbers:

Bismuth ferrite (BiFeO$_3$) is a perovskite ferroelectric with antiferromagnetic order [1, 2] at room temperature. It belongs to a class of material known as multiferroics [2, 3], which have great potentials in industrial applications. Bulk BiFeO$_3$ single crystals have a Curie temperature of $T_C \approx 1103$ K [4–6], and a Néel temperature of $T_N \approx 643$ K [7, 8]. The average crystal structure of BiFeO$_3$ bulk single crystals is a rhombohedrally distorted perovskite [1, 4, 7, 9–12], which belongs to the space group R3C (or $C_6^{3V}$). The rhombohedral unit cell parameters are $a = 3.96$ Å and $\alpha = 89.4°$. Along the [111] direction, there is a three-fold rotation, about which the Bi$^{3+}$ and Fe$^{3+}$ cations are displaced from their centro-symmetric positions. This distortion is polar and results in a polarization of $\mathbf{P} \sim 0.061$ C/m$^2$ oriented along [111]. The magnetic moment is provided by the transition metal cation Fe$^{3+}$. Spins in neighboring atoms are antiparallel [2, 9], resulting in an antiferromagnetic order propagating along along [111] direction.

Recently, epitaxial thin-films of BiFeO$_3$ have been grown on (001) SrTiO$_3$ [13]. Hetero-epitaxy induces significant and important structural changes. The lattice parameters of the epitaxial thin films were found to be different than the bulk rhombohedral ones. The epitaxial layers were reported to have different out-of- and in-plane lattice parameters of 4.005 Å and 3.935 Å, respectively. In addition to the structural changes, the (001) epitaxial BiFeO$_3$ thin films [13] have a dramatically increased spontaneous polarization of $\mathbf{P} \sim 0.6$ C/m$^2$, which is about twenty times larger than that of a bulk crystal projected onto the same orientation. Preliminary x-ray diffraction measurements [14] have been carried out to study the structural changes. Mesh scans were performed around the (101), (202), and (111) Bragg peaks. These results were indicative of a monoclinic structure. But quantitative interpretations were difficult since the fixed (001) reflection was not in the scattering zone, and it was therefore hard to determine the angles between different crystallographic axes. In this paper, we report an improved x-ray diffraction study on the room temperature structures of the BiFeO$_3$ thin-film. It was found that the room temperature structure of the thin film is a slightly modified $M_A$ type monoclinic structure. The diffraction measurements were performed in scattering zones that contain the fixed (001) reflection, and can therefore be used to directly obtain the monoclinic angle(s) of different domains. We will describe first the technique and then results in details below.

Usually powder diffraction measurements are more definitive in determining atomic positions and crystal symmetry. It is nevertheless not applicable in this case. Single crystal diffraction results are very often hard to interpret because of domain effects. Low symmetry phases can have twins or domains with different orientations that lead to complicated splitting patterns of Bragg peaks. The problem is greatly simplified if one crystallographic axis is fixed, and splittings of Bragg reflections are only due to the change of axes orthogonal to the fixed one.

In practice, an electric field along the [001] direction can help to fix the $c$-axis of a ferroelectric perovskite single crystal in space. Similarly, a epitaxial thin-film grown on the (001) substrate will also have its $c$-axis fixed to the normal direction of the substrate surface. When measuring other reflections in a scattering zone containing the fixed (001) reflection, it is possible to determine not only the lengths, but also the orientations of other reflections and axes in different domains/twins, with respect to the fixed (001) reflection, and therefore the crystal structure. Note that it is very important to have the fixed reflection in the scattering zone to have a fixed reference point. The scattering zone is the plane defined by the incident and out-going x-ray beams. For most x-ray diffractometers, this is the plane where the diffraction angle $2\theta$ and sample rotation angle $\omega$ (in some cases, also called $\theta$) are defined. Most four-circle x-ray diffractometers have the ability to go to any point in the reciprocal space permitted by geometry, with the combination of adjusting $2\theta$, $\omega$, $\chi$, and $\phi$ - the last two are usually out-of-plane sample rotation angles. However, the out-of-plane resolution is typically much worse than the in-plane resolution. So most measurements are carried out in the scattering zone. For example, measuring the splitting of (100) or (110) Bragg peaks in the (H0L) or (HHL) zones (see Fig. 1) is often used to obtain the monoclinic angle and lattice parameters in the $M_C$ or $M_A$ phase [15]. Because the (001) reflection does not split, the splitting of (100) or (110) peaks in the corresponding zones can be used to directly obtain the angle between $a_M^*$ (along (100) for $M_C$, (110) for $M_A$) and $c^*$.

We have grown phase-pure BiFeO$_3$ thin films of 2000 Å thickness by pulsed laser deposition (PLD) onto (001) single crystal SrTiO$_3$ substrates. The conducting perovskite oxide electrodes, SrRuO$_3$ [16], was chosen as the bottom electrode due to the closest lattice mismatch with the BiFeO$_3$ structure. Films of SrRuO$_3$ of 500 Å were deposited at 873 K in an oxygen ambient of 100 mTorr; and followed by the BiFeO$_3$ film, deposited at 943 K in an oxygen ambient of 20 mTorr at a growth rate of 0.7 Å/sec. Chemical analysis was carried out by scanning electron microscopy (SEM) x-ray microanalysis, indicating a cation stoichiometry in the BiFeO$_3$ films of $\sim 1:1$. The x-ray diffraction measurements were performed at the National Synchrotron Light Source (NSLS) using beam line X22A. The x-ray energy was 10.2 keV, which can not penetrate the substrate. Therefore only reflective geometry can be used. The out-of-plane lattice parameter of the BiFeO$_3$ thin film was measured before to be $c \approx 3.997$ Å at room temperature. We have used a pseudo-cubic lattice coordinate system with one reciprocal lattice unit $a^* = 2\pi/3.997 = 1.572$ Å$^{-1}$ to describe our results in this report.

Our measurements are described schematically in Fig. 2. With the assumption that the thin film has a possible monoclinic structure, we would like to determine the monoclinic angle by measuring the splitting of (100) (for $M_C$ structure) or (110) (for $M_A$ structure) reflection in the (H0L) or (HHL) zone. However, these reflections can not be reached directly since the substrate will be blocking the x-ray beam. An alternative is to measure other Bragg peaks that are combinations of (100)/(110) with the fixed (001), for example, the (101) reflection. Unfortunately, neither can the (101) reflection be reached considering our sample geometry (Fig. 2 (c)). The (011) reflection, on the other hand, can be reached by rotating the sample by 45° with respect to the $a$-axis. Fig. 2 (d) shows the measurement of the (011) reflection in the (HKK) zone, which is defined by the [100] and [011] vectors. This is the scattering zone used in previous measurements [14], and probably commonly used for studying diffractions from thin films since many Bragg peaks can be probed in this zone. However, the results obtained in this zone are indefinite, because the fixed (001) reflection is not in the zone. Splittings of the (011) peak measured in this zone can only give information on different $d$-spacings of (011) reflections from different domains, but nothing conclusive on the monoclinic angle.

More definitive results can be obtained by measuring the (103) or (113) reflections in the (H0L) or (HHL) zone, as shown in Fig. 2 (e) and (f). First, by going to reflections closer to the $c$-axis, we have been able to move the substrate out of the way of both the incident and out-going x-ray beams. In addition, both scattering zones contain the fixed (001) reflection ($\vec{c}^*$-axis), making it possible for us to determine the monoclinic angles (angles between different crystallographic axes) from splittings of Bragg reflections in the zone.

In Fig. 3 (a), a mesh scan around the (002) Bragg peak in the (H0L) zone is plotted. The (002) reflection is a single peak at the center of the plot, confirming that the $\vec{c}^*$-axis is fixed in both orientation and length. The weak vertical trail and the small bright spot above the main peak are tails from the (002) reflection of the substrate, which has a smaller lattice parameter and therefore a larger **Q**.

In Fig. 3 (b) and (c), mesh scans around the ($\bar{1}\bar{1}3$) and (113) reflections in the (HHL) zone are plotted. Both reflections split into two peaks, (-1.023,-1.023,3.0) and (-1.006,-1.006, 2.98) around ($\bar{1}\bar{1}3$); (1.023,1.023,3.0) and (1.006,1.006,2.98) around (113). Since $c^*$ is fixed and there can be no splittings of (00L) peaks, these splittings can be mapped to splittings of the ($\bar{1}\bar{1}0$) and (110) reflections, i.e., ($\bar{1}\bar{1}0$) reflection splits into (-1.023,-1.023,0) and (-1.006,-1.006,-0.02); (110) reflection splits into (1.023,1.023,0) and (1.006,1.006,-0.02). We can now try to identify these peaks with different domains. As required by symmetry, a (hkl) reflection must be accompanied with a ($\bar{h}\bar{k}l$) reflection. However, we now have a peak at (1.006,1.006,-0.02), but not at (-1.006,-1.006,0.02); a peak at (-1.006,-1.006,-0.02), but not at (1.006,1.006,0.02). This suggests that the domain with [110] tilted up (against $c^*$), was observed around ($\bar{1}\bar{1}3$), showing a peak at (-1.006,-1.006,2.98), but not around (113); the domain with [110] tilted down, was observed around (113), showing a peak at (1.006,1.006,2.98), but not around ($\bar{1}\bar{1}3$); the domain with [110] perpendicular to $c^*$ is observed both at (113) and ($\bar{1}\bar{1}3$). The epitaxial strain in the thin film and local inhomogeneities may contribute to the fact that not all types of domains are observed in a particular region around certain Bragg peaks.

After taking these missing domains into consideration, it is easy to see that the splitting patterns are consistent with a $M_A$ type structure, where the new monoclinic unit cell is doubled and rotated $45°$ in the $a - b$ plane with respect to the primitive pseudocubic one. The (110) reflection should split into three peaks in the (HHL) zone, including one domain with $a_{M_A}$ tilted up, another with $a_{M_A}$ tilted down, and a $b_{M_A}$ domain. The monoclinic angle $\beta$ can be derived from measuring the angle between the tilted (110) reflection to the $c^*$ axis. The lattice parameters $a_{M_A}$ and $b_{M_A}$ can be calculated from the lengths of the (110) vectors. Based on our results, $\beta = 89.2°, a_{M_A}/\sqrt{2} = 3.907$ Å, $b_{M_A}/\sqrt{2} = 3.973$ Å, and $c = 3.997$ Å, for this modified monoclinic phase.

In summary, it is therefore shown that the room temperature structure in the BiFeO$_3$ thin film is a low symmetry phase close to the $M_A$ type monoclinic structure. Our results demonstrate a change in phase stability from rhombohedral in bulk single crystals, to a modified monoclinic structure in epitaxial thin-films. When comparing the in-plane lattice parameter to that of the bulk, it is more useful to compare $d_{100}$ to $a_{bulk} = 3.96$ Å. $d_{100} = \sqrt{a_{M_A}^2 + b_{M_A}^2}/2 = 3.940$ Å, smaller than $a_{bulk}$, but larger than $a_{SrTiO_3} = 3.89$ Å, indicating a clamping effect from the substrate to force the growth of the film to match the substrate lattice. The out-of-plane lattice parameter $c = 3.997$ Å is unique to the film. In addition, the monoclinic angle remains quite close to the rhombohedral angle of the bulk crystal. This structural difference may play a significant role in enhancing the saturation magnetization of (001) films, relative to the oriented crystals, as recently observed [17].

We would like to thank J. Hill, B. Noheda, B. Ocko, and Z. Zhong for stimulating discussions. Financial support from the U.S. Department of Energy under contract No. DE-AC02-98CH10886 and the Office of Naval Research under grants N000140210340, N000140210126, and MURI

N0000140110761 is also gratefully acknowledged.

[a] Permanent address: Institute for Material Research, Tohoku University, Sendai 980-8577, Japan

FIG. 1: Schematics of (100) and (110) reflections, in the (H0L) and (HHL) scattering zones, respectively. (a) and (b) show splittings for the $M_A$ phase; (c) and (d) show splittings for the $M_C$ phase.

FIG. 2: Schematics of the x-ray scattering measurements of different reflections in different scattering zones. The red solid lines in (b) - (f) are the incident and out-going x-ray beams. The purple lines in (c) - (f) are the normal direction of the (001) surface ($c^*$). (a) The original state of the sample without any rotation. (b) Measuring the (002) reflection in the (H0L) zone. (c) Measuring the (101) reflection in the (H0L) zone. Here the incident x-ray beam is blocked by the sample. (d) Measuring the (011) reflection in the (HKK) zone, defined by [100] ($a^*$) and [011] vectors. Here the sample is rotated 45° around the $a$-axis first to put the [011] vector in the scattering zone. (e) Measuring (103) peak in the (H0L) zone. (f) Measuring (113) peak in the (HHL) zone, defined by the [110] and [001] ($c^*$) vectors. Here the sample is rotated 45° around the $c$-axis first to put the [110] vector in the scattering plane.

FIG. 3: X-ray diffraction measurements around the (002), ($\bar{1}\bar{1}3$), and (113) Bragg reflections, performed in the (H0L), (HHL), and (HHL) scattering planes, shown in (a), (b), and (c), respectively.

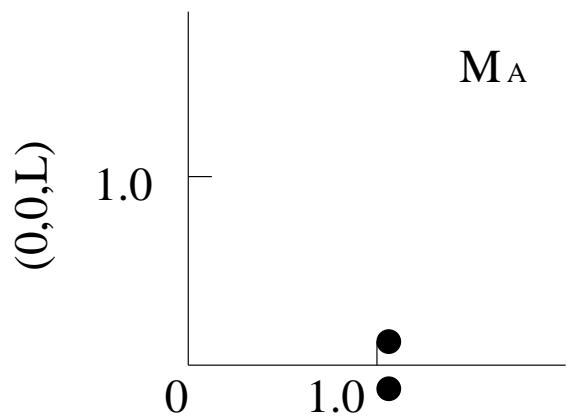 (a)

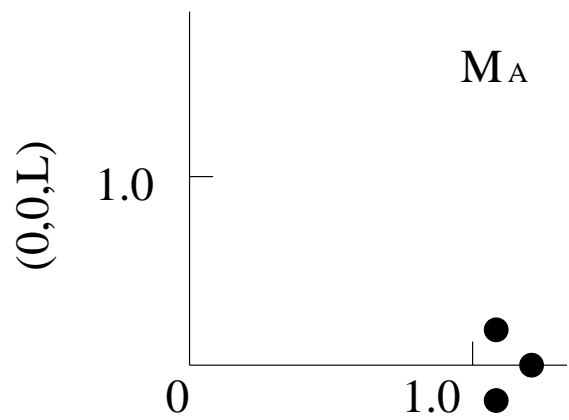 (b)

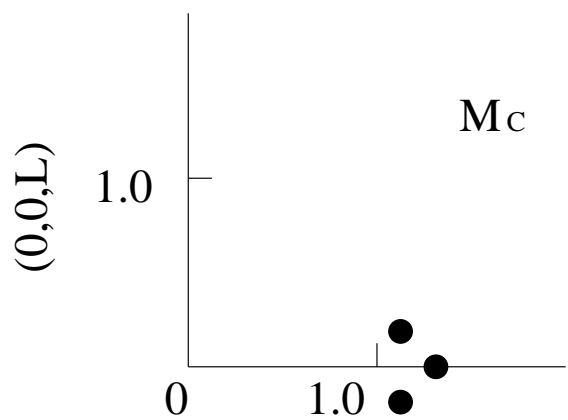 (c)

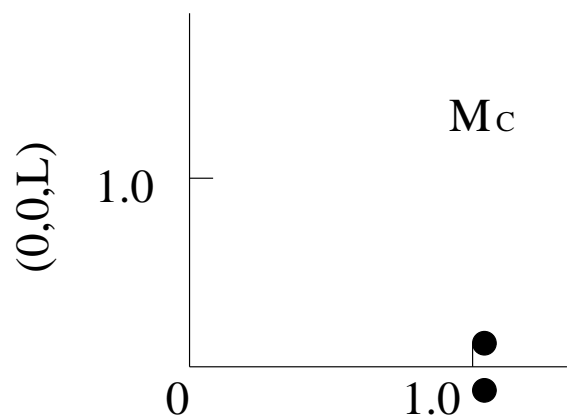 (d)

FIG. 1: Xu *et al.*

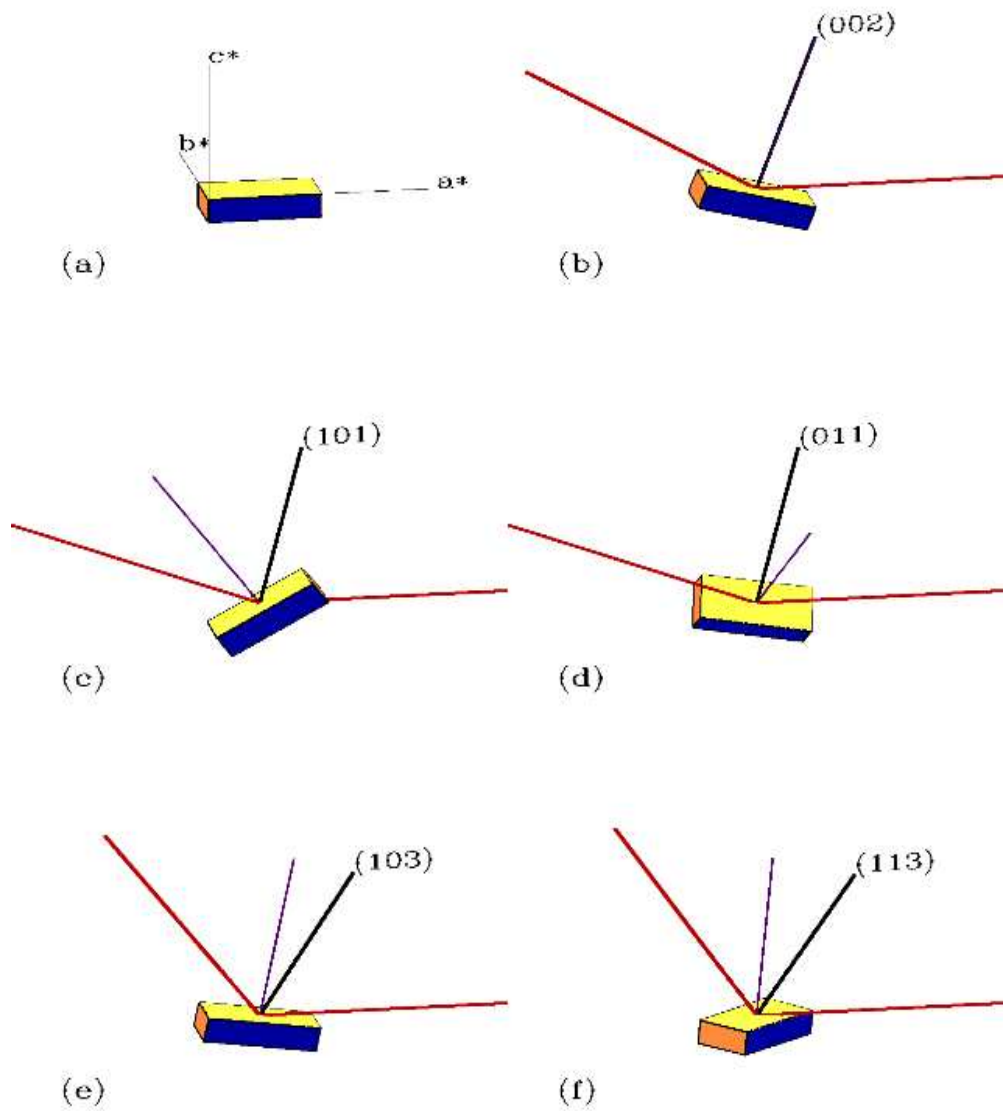

FIG. 2: Xu *et al.*

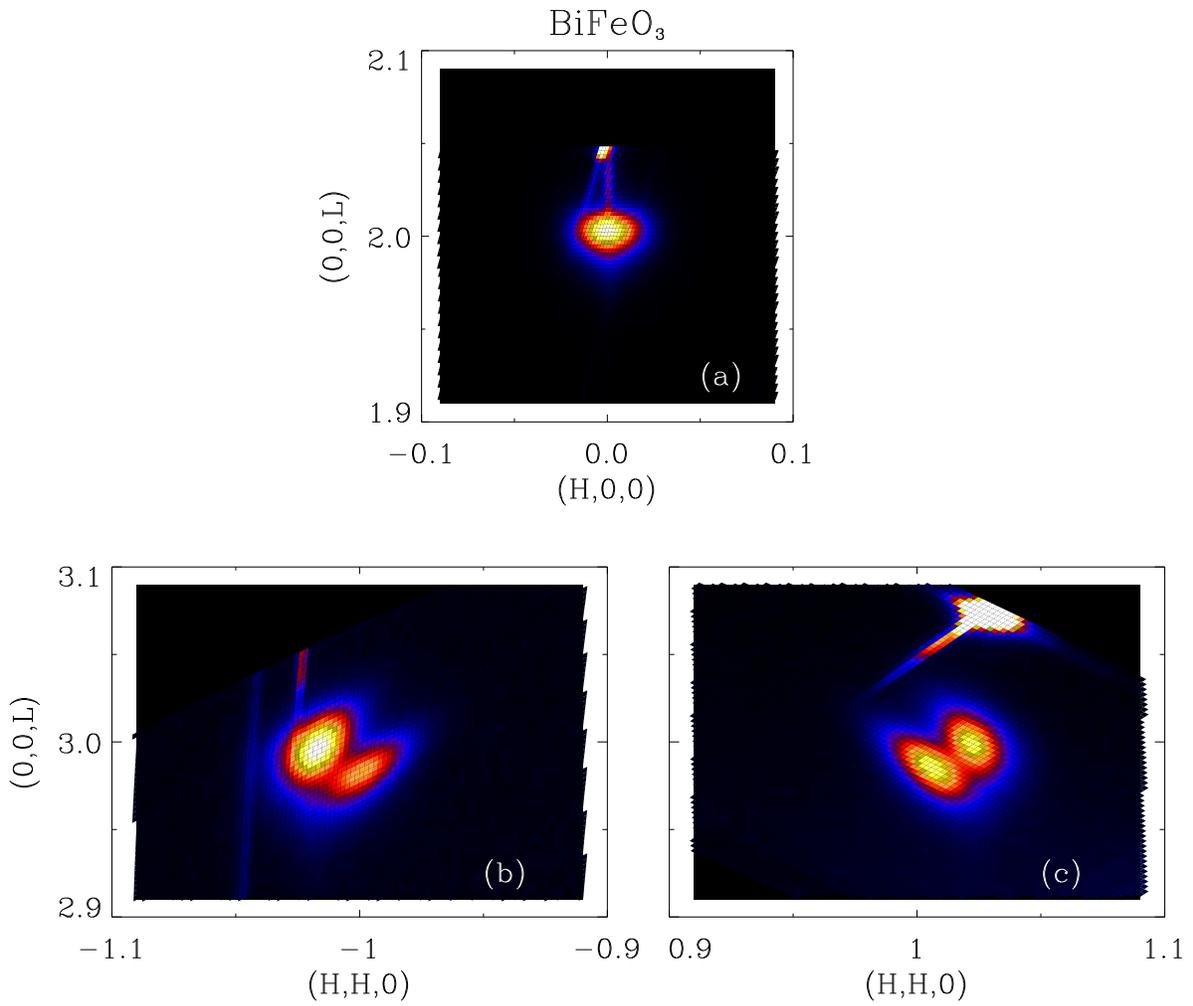

FIG. 3: Xu *et al.*